\documentclass[aps,prl,reprint,superscriptaddress]{revtex4-2}
\usepackage{amsmath}
\usepackage{amssymb}
\usepackage{amsfonts}
\usepackage{bbm}
\usepackage{bm}
\usepackage{graphicx}
\usepackage{color}
\usepackage{braket}
\usepackage{dcolumn}
\usepackage{MnSymbol}
\usepackage{times}
\usepackage[qm]{qcircuit}
\usepackage[colorlinks=true,linkcolor=blue,citecolor=blue,urlcolor=blue]{hyperref}
\usepackage{listings}
\usepackage{matlab-prettifier}

\newcommand*{\id}{\mathbbm{1}}
\newcommand*{\Tr}{\textrm{Tr}}
\newcommand{\inp}[1]{\langle{#1}\rangle}
\newcommand*{\tr}{\mathrm{tr}}

\def\map{\mathcal}

\usepackage{color}
\definecolor{dred}{rgb}{.8,0.2,.2}
\definecolor{ddred}{rgb}{.8,0.5,.5}
\definecolor{dblue}{rgb}{.2,0.2,.8}
\definecolor{dgreen}{rgb}{.2,0.5,.2}

\usepackage{amsthm,amsmath,amssymb}
\usepackage{mathrsfs}

\newcommand*{\physus}{Department of Physics, Southern University of Science and Technology, Shenzhen 518055, China}
\newcommand*{\inssus}{Shenzhen Institute for Quantum Science and Engineering, Southern University of Science and Technology, Shenzhen 518055, China}

\begin{document}
\title{Quantum causal inference via scattering circuits in NMR}

\author{Hongfeng Liu}
\thanks{These authors contributed equally to this work.}
\affiliation{\physus}

\author{Xiangjing Liu}
\thanks{These authors contributed equally to this work.}
\affiliation{CNRS@CREATE, 1 Create Way, 08-01 Create Tower, Singapore 138602, Singapore }
\affiliation{MajuLab, CNRS-UCA-SU-NUS-NTU International Joint Research Unit, Singapore}
\affiliation{Centre for Quantum Technologies, National University of Singapore, Singapore 117543, Singapore}
\affiliation{\physus}

\author{Qian Chen}
\affiliation{Univ Lyon, Inria, ENS Lyon, UCBL, LIP, F-69342, Lyon Cedex 07, France}

\author{Yixian Qiu}
\affiliation{Centre for Quantum Technologies, National University of Singapore, Singapore 117543, Singapore}

\author{Vlatko Vedral}
\affiliation{Clarendon Laboratory, University of Oxford, Parks Road, Oxford OX1 3PU, United Kingdom}

\author{Xinfang Nie}
\email{niexf@sustech.edu.cn}
\affiliation{\physus}
\affiliation{Quantum Science Center of Guangdong-Hong Kong-Macao Greater Bay Area, Shenzhen 518045, China}

\author{Oscar Dahlsten}
\email{oscar.dahlsten@cityu.edu.hk}
\affiliation{Department of Physics, City University of Hong Kong, Tat Chee Avenue, Kowloon, Hong Kong SAR, China}
\affiliation{\physus}
\affiliation{\inssus}
\affiliation{Institute of Nanoscience and Applications, Southern University of Science and Technology, Shenzhen 518055, China}

\author{Dawei Lu}
\email{ludw@sustech.edu.cn}
\affiliation{\physus}
\affiliation{\inssus}
\affiliation{Quantum Science Center of Guangdong-Hong Kong-Macao Greater Bay Area, Shenzhen 518045, China}
\affiliation{International Quantum Academy, Shenzhen 518055, China}

\date{\today}

\begin{abstract}
We report NMR scattering circuit experiments that reveal causal structure. The scattering circuit involves interacting a probe qubit with the system of interest and finally measuring the probe qubit. The scattering circuit thereby implements a coarse-grained projective measurement. Causal structure refers to which events influence others and in the quantum case corresponds to different quantum circuit structures. In classical scenarios, intervention is commonly used to infer causal structure. In this quantum scenario of a bipartite system at two times, we demonstrate via scattering circuit experiments that coarse-grained measurements alone suffice for determining the causal structure. The experiment is undertaken by manipulating the nuclear spins of four Carbon-13 atoms in crotonic acid. The data analysis determines the compatibility of the data with given causal structures via representing the data as a pseudo density matrix (PDM) and analysing properties of the PDM. We demonstrate the successful identification of the causal structure for partial swap channels and fully decohering channels. 
\end{abstract}

\maketitle


\section{Introduction} 
\begin{figure}[t]
    \includegraphics[width=0.85\linewidth]{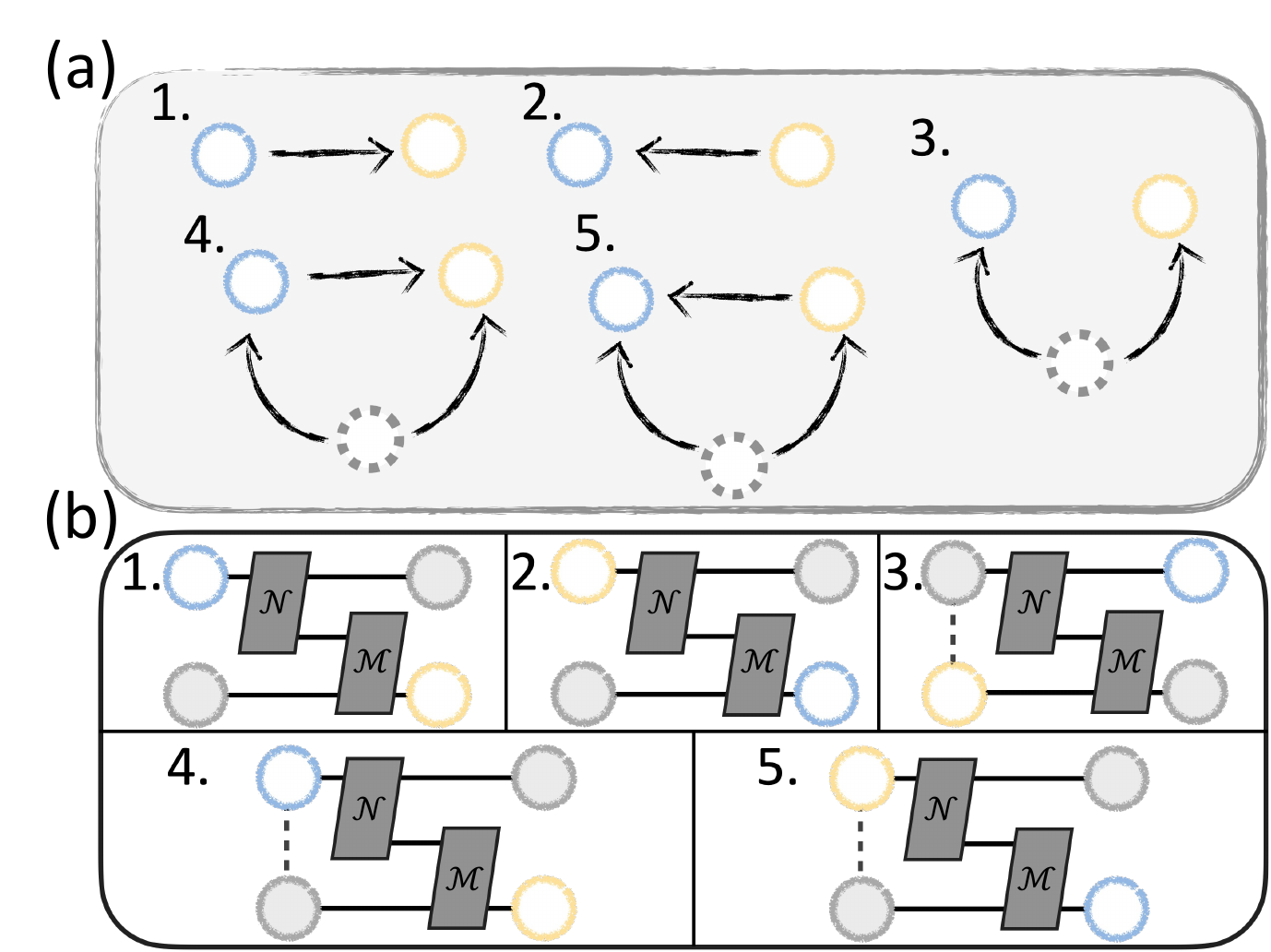}
    \caption{{\bf Causal structures and corresponding quantum circuits.} (a) Five causal structures that are distinct by Reichenbach’s principle: 1) $A$ (blue ball) directly influences $B$ (yellow ball); 2) $B$ directly influences $A$; 3) $A$ and $B$ are influenced by a common cause (represented by a dashed grey ball), indicating correlations in the initial state; 4) a combination of cases 1 and 3; 5) a combination of cases 2 and 3. (b) Quantum circuits reflecting the structures of (a). $N$ and $M$ are unitary interactions. Dashed lines indicate correlations in the initial state. }
\label{fig:Fig1} 
\end{figure}
Causal structure inference in quantum systems is actively investigated~\cite{leggett1985quantum, oreshkov2012quantum,brukner2014quantum,fitzsimons2015quantum,chaves2015information,barrett2021cyclic,barrett2019quantum,hardy2005probability,costa2016quantum,allen2017quantum,PhysRevA.79.052110,PhysRevLett.129.230604,PRXQuantum.4.020334,PhysRevX.11.021043,song2023causal,wolfe2020quantifying}.  
Causal inference refers to understanding which events influence which~\cite{reichenbach1956direction, pearl2009causality}. A natural approach is to reset the state of a subsystem at a given time point and investigate whether that influences later events~\cite{ried2015quantum,bai2022quantum,chiribella2019quantum,maclean2017quantum,chaves2018quantum,agresti2022experimental, gachechiladze2020quantifying,nery2018quantum,agresti2020experimental}. Such resets are not always necessary to determine the causal structure and measurements can suffice~\cite{ried2015quantum, liu2023inferring,fitzsimons2015quantum}.

Collectively, these encouraging findings indicate the feasibility of expanding causal inference into the quantum realm, hinting at a potentially notable divergence from the classical scenario wherein measurements alone could be adequate, at least for some cases. We therefore here push the boundaries causal inference based on measurements.  We present an experimental validation of a minimally invasive quantum causal inference protocol, expanding on the results of the accompanying Ref.~\cite{LiuShort}. 

The experiments were conducted using an NMR platform. As illustrated in FIG.\ref{fig:Fig1}, the observer collects data by observing two quantum systems, denoted $A$ and $B$, at two distinct time points. The objective is to discern the causal relationships within the process that generates the observed data. The proposed causal inference method does not involve reset-type interventions but relies solely on coarse-grained projective measurements. These light touch measurements are carried out using a technique known as scattering circuits~\cite{miquel2002interpretation,PhysRevA.105.L030402}. These are unitary interactions between a probe qubit and a system of interest. A final fine-grained projective measurement on the probe qubit implements a coarse-grained measurement on the multiqubit system of interest. 

In our data analysis to determine the causal structure from the measurement data, the initial step involves identifying negativity within specific reduced states of the pseudo-density matrix (PDM) representing the experimental data over two times and two systems~\cite{fitzsimons2015quantum,liu2023quantum}. Subsequently, the time asymmetry of the PDM is evaluated. 

We demonstrate successful causal inference for two types of channels: partial swap channels and a fully decohering channel. The latter experiment demonstrates that causal inference via measurements is possible even without coherence in the channel.

\section{Theoretical methods}
Our protocol builds on a quantum space-time states formalism called the Pseudo-Density Matrix (PDM) formalism~\cite{fitzsimons2015quantum,liu2024unification,zhang2020different,liu2023inferring}. This formalism generalizes the conventional density matrix by assigning a Hilbert space to each instant in time, as described in the reference~\cite{fitzsimons2015quantum}. The central focus of our work will be the 2-time n-qubit PDM, which we will use as the primary tool to infer quantum causal structure. This 2-time n-qubit PDM is formally defined as follows:
\begin{equation}\label{eq: defPDM}
    R_{12}=\frac{1}{2^{2n}}\sum^{4^n-1}_{i_1, i_2=0} \langle \sigma_{i_1},  {\sigma}_{i_2} \rangle \, {\sigma}_{i_1} \otimes   {\sigma}_{i_2},
\end{equation}
where ${\sigma}_{i_\alpha}\in \{\mathbbm{1},\sigma_x,\sigma_y,\sigma_z\}^{\otimes n}$ is an $n$-qubit Pauli matrix at times $t_\alpha, \alpha=1,2$ and $\langle \sigma_{i_1},  \sigma_{i_2} \rangle$ denotes the expectation value of the product of observables $\sigma_{i_1}$ at $t_1$ and $\sigma_{i_2}$ at  $t_2$. 
For $\map {N}$ being the channel between the two times there is a corresponding Choi–Jamio{\l}kowski (CJ) matrix~\cite{choi1975completely,jamiolkowski1972linear}
\begin{align}
\label{eq:CJ}
    M=\sum^{2^n-1}_{i,j=0} \ket{i} \bra{j} \otimes \map N (\ket{j} \bra{i}).
\end{align} 
There is a neat closed form of the PDM in terms of the CJ matrix. If the implementation of each $\sigma_{i_\alpha} $ is chosen to be the coarse-grained projectors
\begin{align}
\label{eq:coarseprojectors}
    \left\{ \Pi^{i_\alpha}_+= \frac{\id + \sigma_{i_\alpha} }{2}, \, \Pi^{i_\alpha}_-= \frac{\id - \sigma_{i_\alpha} }{2}  \right\},
\end{align}
the 2-time multiple qubit PDM has been proven to be~\cite{liu2023quantum}
\begin{align} \label{eq: ClosedForm}
    R_{12}=\frac{1}{2} (\rho \, M_{12} + M_{12} \, \rho),
\end{align}
where $\rho=\rho_1 \otimes \id_2$. Eq.~\ref{eq: ClosedForm} generalizes an expression for PDMs of two Hilbert spaces (two times or two locations)~\cite{fitzsimons2015quantum}.

The PDM has several similarities with the standard density matrix~\cite{fitzsimons2015quantum,liu2023quantum}. The PDM is Hermitian with unit trace. The partial trace is well-defined in the natural way. The standard density matrix is recovered if the Hilbert spaces for all but one time are traced out. 

A key difference between the PDM and the standard density matrix, which is useful for our purposes, is the possibility of negative eigenvalues. These negative eigenvalues are inconsistent with spatial correlations since the standard density matrix is positive.  The negativity must thus serve as an indicator of temporal correlation and causal influence. A natural measure for quantifying the negative eigenvalues of a matrix operator $O$ is defined as 
\begin{equation}\label{Eq5}
    f(O):=\Tr \sqrt{OO^\dag} - \Tr \, O .
\end{equation}
The measure $f$ quantifies the absolute value of the sum of the negative eigenvalues of $O$. Under this measure, $f(O)>0$ when the matrix operator $O$ has negative eigenvalues, and $f(O)=0$ when $O$ is positive semidefinite. Therefore, it is useful to know whether  $f(R)>0$, as that indicates a temporal scenario.

To determine the time ordering in the temporal scenario ($f(R)>0$), we compare the PDM with a corresponding time-reversed PDM following Refs.\cite{liu2023inferring,liu2023quantum}. The time-reversed PDM is defined by
\begin{equation}\label{eq:reversePDM}
R_{21}:= S \, R_{12} \, S^\dag, 
\end{equation}
where $S$ denotes the operator that swaps the systems at times $t_1$ and $t_2$. The time-reversed PDM has a similar closed-form expression to Eq.~\eqref{eq: ClosedForm}:
\begin{align}
R_{21} =\frac{1}{2}( \pi \, M_{21} +M_{21} \, \pi ),
\end{align}
where $\pi:= (\Tr_{1} R_{12}) \otimes \id_1 $ and $ M_{21}$ is the CJ matrix (Eq.\ref{eq:CJ}) of the time-reversed process.  

The CJ matrices $M_{12}$ and ${M}_{21}$ can be extracted uniquely from the PDMs $R_{12}$ and $R_{21}$ respectively, when the initial state for the given time ordering is full rank~\cite{liu2023inferring,liu2023quantum}. For rank-deficient initial states, there are infinitely many CJ matrices but an algorithm along with the code in Sec.\ref{sec:SDP} gives the least negative Choi matrices from which the CJ matrix is directly obtained via partial transpose.

\begin{table}
\caption{Summary of protocol}
\label{tab:protocol}
\begin{tabular}{|c|c|c|c|}
  \hline
  $f(R_{AB})$ & $f(M^T_{AB})$ & $f(M^T_{BA})$ & $R_{AB}$ is compatible with  \\
  \hline
 0 & Any & Any & Common Cause \\
  \hline
  $>0$ & 0 &   $>0$ & $ A \rightarrow B $ \\
  \hline
  $>0$ &   $>0$ & 0  &   $B \rightarrow A $ \\
  \hline
    $>0$ & 0 & 0 &  $ A \rightarrow B $ or  $ B \rightarrow A $\\
  \hline
      $>0$ & $>0$ & $>0$ & Mixture\\
  \hline
\end{tabular}
\vspace{.2cm}
\end{table}

The CJ matrices $M_{12}$ and ${M}_{21}$ can be used to determine the time ordering. A process being completely positive (CP) is equivalent to its Choi matrix being positive~\cite{choi1975completely,jamiolkowski1972linear}. The Choi operators here are $M^{T}_{12} $ and $M^{T}_{21}$, where $T$ denotes the transpose on the initial quantum system. The dynamics of a quantum system can be modeled by a CP map when the system is initially uncorrelated from the environment. Therefore, if there is no common cause (no initial correlation) the Choi matrix of the forward process is positive semidefinite, i.e., $f(M^T_{12})=0$. After the dynamics, the system is in general correlated with the environment. In this case, a CP map in general cannot describe the time-reversed process. This means that the Choi matrix of the reverse process may have negative eigenvalues, $f(M^{T}_{21})>0$. Therefore, when only one of the two Choi matrices is positive we say there is asymmetry of the temporal quantum correlations, and the time order is revealed.

The time ordering is subtle in other scenarios. Firstly, if both Choi matrices are positive semidefinite, the experiment can in principle align with either direction.  Secondly, if both Choi matrices are not CP ($f(M^{T}_{12})>0$ and $f(M^{T}_{21})>0$) then neither process is CP and there must be a common cause (initial correlations).

The above reasoning implies a protocol that determines the compatibility of the experimental data with the five causal structures of Fig.\ref{fig:Fig1}. In line with causal inference terminology, we say that the data and a causal structure are  {\em compatible} if that structure could have generated experimental data.
The protocol is summarized in table \ref{tab:protocol}.

\section{Quantum scattering circuit for measuring 2-time correlators}
\label{sec: scatteringcircuit}

\begin{figure}
    \includegraphics[width=1\linewidth]{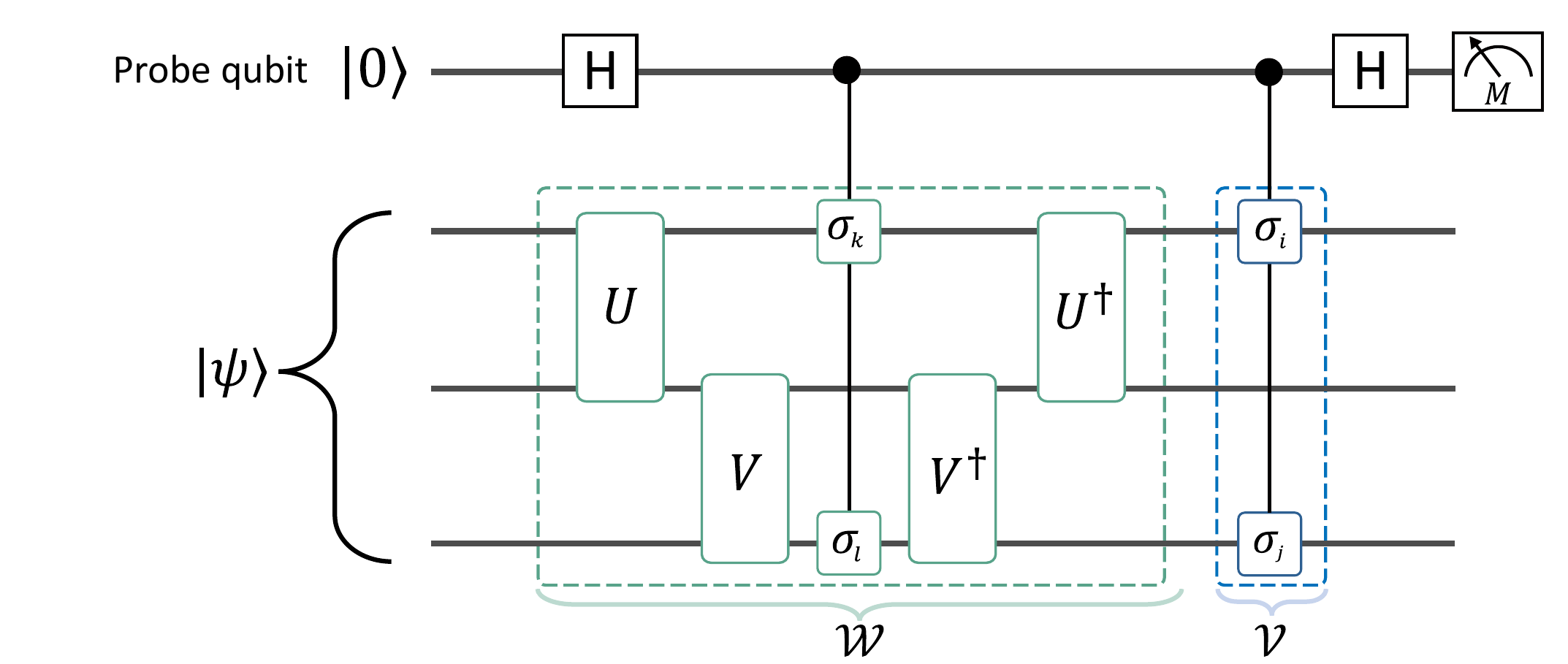}
    \caption{{\bf Scattering quantum circuit used in the experiment.} The circuit  measures the expectation values $\langle \sigma_i\otimes\sigma_j,\sigma_k\otimes\sigma_l\rangle$, under a unitary evolution. In this example, the unitary evolution is the partial swap $e^{-iS\theta}$ where $S$=SWAP and $\theta \in [0,2\pi]$. The two controlled Pauli measurements, one inside $\mathcal{W}$ and one constituting $\mathcal{V}$ determine the measurement choice. }
\label{mainlongfig: f2} 
\end{figure}

The 2-time PDM employed to infer the quantum causal structure is constructed from the two-time correlators and the corresponding Pauli operators. To derive a closed-form of the 2-time PDM, the coarse-grained measurements are used to obtain the two-time correlators. This section shows that one can obtain the same values of the two-time correlators by the quantum scattering circuit shown in Fig.~\ref{mainlongfig: f2}. The demonstration is divided into the following two steps.

Firstly, we show that the expectation value of $\sigma_z$ on the output of the probe qubit, $\langle \sigma_z \rangle_{ \text{probe}}$, equals the two-time correlator. In Fig.~\ref{mainlongfig: f2}, the probe qubit is in the state of $\ket{0}$ and the other three qubits are in the state of $\ket{\psi}$. The overall state undergoes the following evolution:
\begin{align}
   \ket{0}\ket{\psi} &\to \ket{+} \ket{\psi} \to \frac{1}{\sqrt{2}}(\ket{0}\ket{\psi}+ \ket{1} \mathcal{W}\ket{\psi} \nonumber\\
   &\to  
   \frac{1}{\sqrt{2}}(\ket{0}\ket{\psi}+ \ket{1} \mathcal{V}\mathcal{W}\ket{\psi} ) \nonumber\\
   &\to \frac{1}{2}(\ket{0}(\mathbb{I}+\mathcal{V}\mathcal{W})\ket{\psi} +\ket{1}(\mathbb{I}-\mathcal{V}\mathcal{W})\ket{\psi} ) ,
\end{align}
where $\mathcal{W}= U^\dag V^\dag 
 ( \sigma_k \otimes \sigma_l )   V U$ and $\mathcal{V}=\sigma_i \otimes \sigma_j$. Therefore, the value of $\langle \sigma_z \rangle_{ \text{probe}}$  at the end is given by
\begin{align}\label{eq:sigmaz}
    \braket{\sigma_z}_{\text{probe}} =& \frac{1}{4} \bra{\psi}(2\mathbb{I} + \mathcal{V}\mathcal{W}+\mathcal{W}^\dag\mathcal{V}^\dag)\ket{\psi} \nonumber\\
    &-\frac{1}{4} \bra{\psi}(2\mathbb{I} -\mathcal{V}\mathcal{W}-\mathcal{W}^\dag\mathcal{V}^\dag)\ket{\psi} \nonumber\\
    =& \frac{1}{2} \bra{\psi}(\mathcal{V}\mathcal{W}+\mathcal{W}^\dag\mathcal{V}^\dag)\ket{\psi} \nonumber\\
    =& \Tr\left[   (\sigma_i \otimes \sigma_j)VU \rho (\sigma_k \otimes \sigma_l ) V^\dag U^\dag   \right] \nonumber\\
    =& \braket{\sigma_i\otimes\sigma_j, \sigma_k \otimes\sigma_l },
\end{align}
where here  $\rho=\ket{\psi}\bra{\psi}$. (Note that we sometimes omit the tensor product symbol when there is no risk of confusion.)
The generalization of the above calculation to mixed states is straightforward.

Secondly, we show that the quantity $\braket{\sigma_z}_{\text{probe}}$ in Eq.~\eqref{eq:sigmaz} equals the two-time correlators obtained by the coarse-grained measurements of Eq.~\ref{eq:coarseprojectors}. In particular,
\begin{align}
    &\braket{\sigma_i\otimes\sigma_j, \sigma_k \otimes\sigma_l } \nonumber\\
    =& \braket{\Pi^{(i,j)}_+-\Pi^{(i,j)}_-, \Pi^{(k,l)}_+-\Pi^{(k,l)}_-} \nonumber\\
=&\braket{\Pi^{(i,j)}_+, \Pi^{(k,l)}_+}+\braket{\Pi^{(i,j)}_-, \Pi^{(k,l)}_-} \nonumber\\
    &-(\braket{\Pi^{(i,j)}_-, \Pi^{(k,l)}_+}+\braket{\Pi^{(i,j)}_+, \Pi^{(k,l)}_-}).
\end{align}
Here, $\Pi^{(i,j)}_{+}= \Pi^i_+ \otimes \Pi^j_+ +\Pi^i_- \otimes \Pi^j_- $ denotes the coarse-grained projector onto the $+1$ eigenspace of $\sigma_i\otimes\sigma_j$, $\Pi^{(i,j)}_{-}= \Pi^i_- \otimes \Pi^j_+ +\Pi^i_- \otimes \Pi^j_+ $ denotes the coarse-grained projector onto the $-1$ eigenspace of $\sigma_i\otimes\sigma_j$, with a similar definition for $\Pi^{(k,l)}_{\pm}$.

\section{Implementing quantum circuits and measurements via NMR}\label{metho}
\begin{figure}[t]
\centering
\includegraphics[width=1\linewidth]{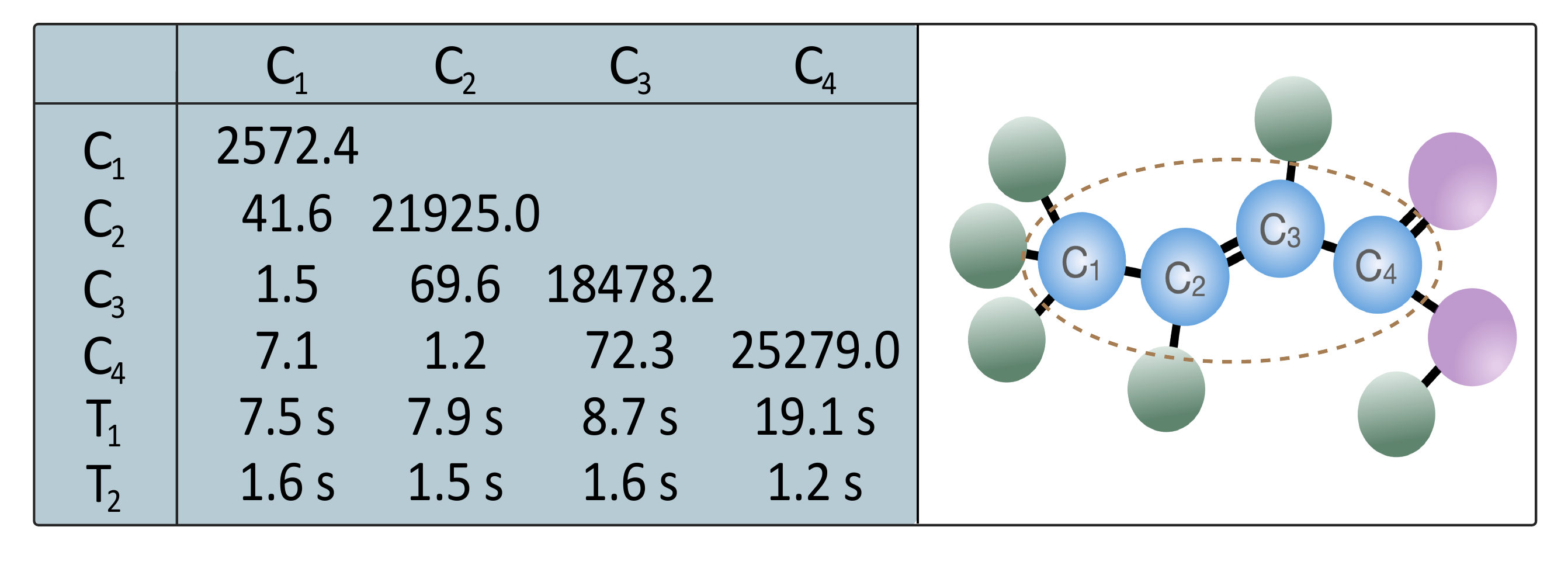}
\caption{{\bf Molecular structure and the NMR parameters of the sample.} The chemical shifts $\nu_i$ and the scalar coupling strengths $J_{ij}$ are given in the diagonal and off-diagonal segments of the accompanying table, respectively. The longitudinal and transverse relaxation times $T_{1,2}$ (in seconds) are shown at the bottom.}
\label{molecular}
\end{figure}

\textit{Characterization.}--Our experiments were conducted on a Bruker 600 MHz spectrometer ~\cite{Xin_2018,RevModPhys.76.1037,JONES201191,KHANEJA2005296,park2016simulation}, using the 
$^{13}$C-labeled crotonic acid molecules dissolved in deuterated acetone ($d_6$-acetone) as a quantum simulator.
The four coupled $^{13}$C spins correspond to the four qubits of the quantum simulator.
The internal dynamics of the system 
is governed by the internal Hamiltonian
\begin{equation}
\label{eq:NMRHamiltonianSuppl}
\mathcal{H}_{\text{int}} = \pi\sum^4_{i=1}\nu_i\sigma^i_z + \pi\sum^4_{i<j,=1}\frac{J_{ij}}{2}\sigma^i_z\sigma^j_z,
\end{equation}
where $\sigma_z^i$ represents the Pauli-Z matrix of the $i$-th spin, $\nu_i$ is the chemical shift of the $i$-th spin and $J_{ij}$ is the J-coupling strength between the $i$-th and $j$-th spins. The specific structure of the four utilized spins, along with their respective parameters, is depicted in Fig.~\ref{molecular}. In this setup, $^{13}$C$_1$ acts as the probe qubit, $^{13}$C$_{2,4}$ constitute the two system qubits (A and C), and $^{13}$C$_3$ is the ancillary qubit to simulate the effect of the channel $\mathcal{M}$.
The system is controlled using a radio-frequency (RF) field, with control Hamiltonian described as 
\begin{equation}
\mathcal{H}_{\text{con}} = - B_1\sum\limits_{i = 1}^4\gamma_i[\cos(\omega_{rf}t+ \phi)\sigma_x^i+\sin(\omega_{rf}t+ \phi)\sigma_y^i],
\end{equation}
where $B_1$, $\omega_{rf}$ and $\phi$ represent the pulse amplitude, frequency and phase of the control pulse, respectively. Single-qubit rotations are achieved by applying RF pulses, while two-qubit gates utilize scalar coupling between different spins combined with the spin echo technique \cite{RevModPhys.76.1037}.

\begin{figure}[b]
   \includegraphics[width=1\linewidth]{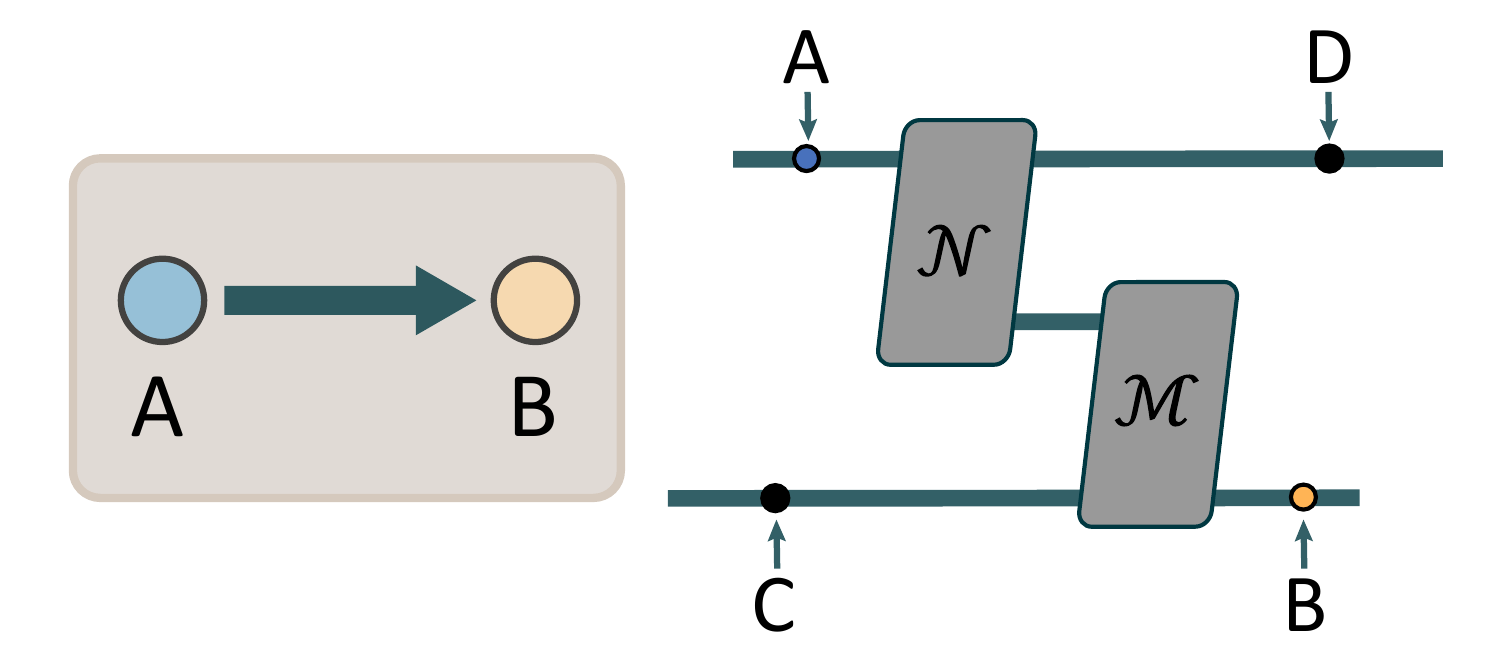}
   \caption{{\bf Causal-effect mechanism and the circuit model.} The diagram on the left shows the causal structure, while the one on the right illustrates the circuit model. In the causal structure, the directed acyclic graph represents quantum systems as nodes and causal influence as directed edges. The circuit model depicts the channel evolution of the quantum system across different time points. The points $A$, $B$, $C$, and $D$ in the circuit denote the quantum systems in spacetime on which measurements are made to construct the associated PDMs.}
\label{mainlongfig: f3} 
\end{figure}

\begin{figure*}[t]
\centering
\includegraphics[width=1\linewidth]{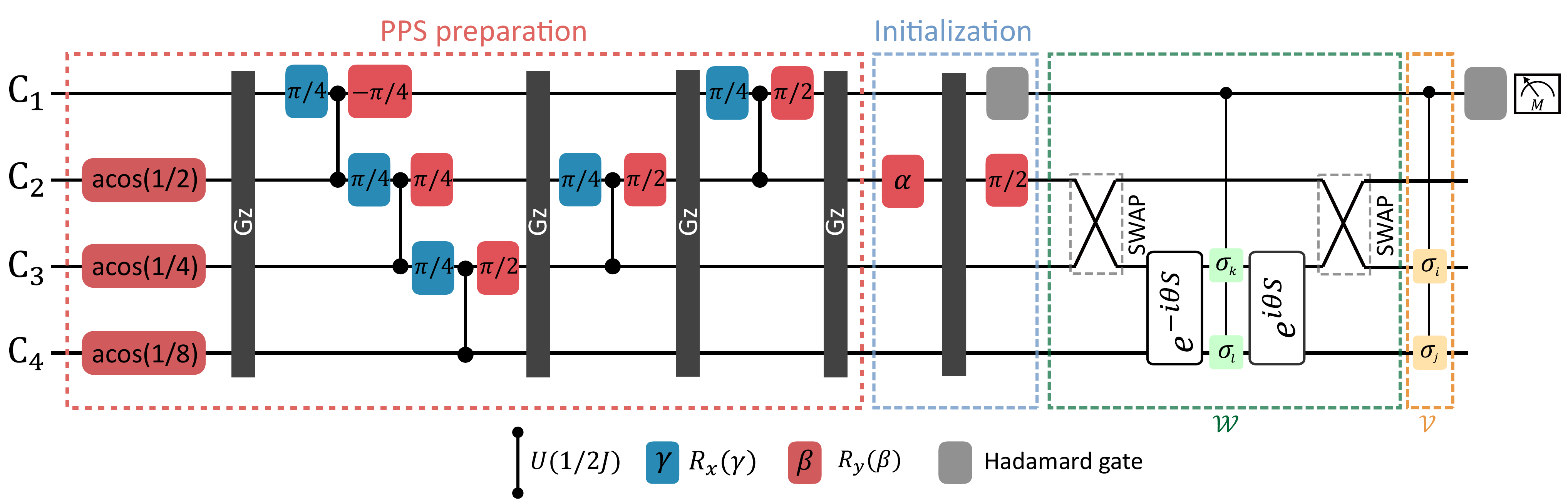}
\caption{
{\bf The quantum circuit for experimental implementation.} 
The operations encased in the red color dashed box are utilized for generating the pseudo-pure state via the spatial averaging method. Following this, the blue dashed box indicates the preparation of the desired initial state of the quantum casual inference experiment. Then the experimental quantum circuit used to measure the expectation values $\langle \sigma_i^A\sigma_j^C,\sigma_k^D\sigma_l^{B}\rangle$.}
\label{circuit}
\end{figure*}

\textit{Pseudo-pure state preparation.}--Starting from an initial thermal equilibrium
state, we first create a pseudo-pure state (PPS).
The thermal equilibrium state of the spin system at room temperature is described by
\begin{equation}
\rho_{\text{eq}} = \frac{1-\epsilon}{16}\mathbb{I} + \epsilon\sum^4_{i=1}\sigma_z^i,
\label{eqstate}
\end{equation}
where $\mathbb{I}$ is the $16 \times 16$ identity matrix, and $\epsilon$, representing polarization, is approximately $10^{-5}$. Several methods exist for system initialization in NMR, including the spatial average method, the selective transition method, the time-spatial method, and the cat-state method. For our experiment, we opted for the spatial average method to initialize the NMR system, employing the pulse sequence depicted in Fig.~\ref{circuit}. In the circuit diagram, the single-qubit rotations are implemented using rf pulses. The two-qubit gates are executed by exploiting the scalar coupling and spin echo technique and optimized utilizing shaped pulses. This pulse sequence transforms the equilibrium state described in Eq.~\eqref{eqstate} into a PPS
\begin{equation}
\rho_{\text{PPS}} = \frac{1-\epsilon'}{16}\mathbb{I} + \epsilon'\op{0000}{0000},
\label{eq:PPS}
\end{equation}
where $\epsilon'$ is the polarization of the PPS. The substantial component of the identity matrix $\mathbb{I}$ remains unchanged under any unitary transformation and is undetectable in NMR experiments, this allows the quantum state to be regarded as the pure state $\op{0000}{0000}$. 

\textit{Initialization.}--In the partial swap channel experiment, we initialize the system to $\rho_0=\ket{+}\bra{+}\otimes\rho_A\otimes\ket{00}\bra{00}$, where $\rho_A=(1-\lambda) \mathbb{I}/2+\lambda\ket{+}\langle+|$. The mixed state $\rho_A$ can be prepared using two single-qubit gates alongside a gradient-field pulse, presented by the circuit depicted in the blue dashed box in Fig.~\ref{circuit}. The first single-qubit rotation, denoted as $R_y(\alpha)$, rotates the qubit to the state $\sqrt{(1+\lambda)/2}\vert0\rangle+\sqrt{(1-\lambda)/2}\vert1\rangle$, where $\alpha = \text{arccos}\sqrt{\langle \sigma_x \rangle^2 + \langle \sigma_y \rangle^2 + \langle \sigma_z \rangle^2}$, 
with $\langle \sigma_i \rangle = \text{Tr}(\rho_A \sigma_i)$. 
Subsequently, a gradient-field pulse is applied to dephase the state, remove the coherence and transform the state into the mixed state, $(1-\lambda)\mathbb{I}/2+\lambda\vert0\rangle\langle0\vert$, achieving the same purity as the target state $\rho_A$. Following this, a second single-qubit rotation $R_y(\pi/2)$ evolves the state into $\rho_A$. Simultaneously, a Hadamard gate is applied to the probe qubit to prepare it in $\vert+\rangle\langle+\vert$.

\section{Experimental implementation of protocol for structure with coherent channel}
We experimentally demonstrate our light-touch quantum causal inference protocol on the NMR platform by inferring a cause-effect mechanism. As shown in the left panel of Fig.~\ref{mainlongfig: f3}, the directed acyclic graph represents the cause-effect mechanism in which only $A$ can influence $B$ and there is no shared cause (initial correlation). 

This causal structure can be modeled by the circuit depicted in the right panel of Fig.~\ref{mainlongfig: f3}. Three systems are involved in the circuit. As we see, a quantum system interacts with an ancillary system via the channel $\mathcal{N}$, and then the ancillary system interacts with another quantum system described by the channel $\mathcal{M}$. Both channels are non-trivial. Further, we label quantum systems at different times as $A, B, C$, and $D$, and there are no correlations between $A$ and $C$.  
To facilitate the experiment, we consider those quantum systems as qubits and set the channels to be  
\begin{align}\label{eq:channels}
   \mathcal{N}(\cdot)=S(\cdot)S^{\dag} \ \text{and} \ \mathcal{M}(\cdot)=e^{-i\theta S}(\cdot)e^{i\theta S},  
\end{align}
 where $S$ denotes the SWAP operator and $\theta$ is a parameter. The initial state is as follows:
 \begin{align}\label{eq:initialAC}
      \rho_{AC}=[ \lambda\ket{+}\bra{+} +(1-\lambda)\mathbb{I}/2 ]\otimes\ket{0}\bra{0}.
 \end{align}
Here, $\lambda$ is the polarization of  $A$.

The quantum circuit is depicted in Fig.~\ref{circuit}. The PPS preparation and initialization segments were described in the experimental setup. We now describe the subsequent parts.

{\em Scattering circuit}.---Our quantum causal inference experiment relies on constructing the PDM $R_{AB}$, which depends on the coarse-grained measurement of the expectation values $\langle \sigma_i^A, \sigma_j^B \rangle$. To minimize the disturbance on the system caused by the measurement process, we introduce the quantum scattering circuit~\cite{miquel2002interpretation,PhysRevA.105.L030402}, as illustrated in Fig.~\ref{mainlongfig: f2}, to obtain the expectation value $\langle \sigma_i^A, \sigma_j^B \rangle$. The circuit is equipped with a probe qubit that interacts with the quantum system. By doing nothing on $C$ and $D$ and then measuring the probe qubit, we have
 \begin{align}
 \langle \sigma_z \rangle_{ \text{probe}}=  \langle \sigma_i^A, \sigma_j^B \rangle :=   \langle \sigma_i^A \id^C, \id^D \sigma_j^B \rangle .
 \end{align}
In this way, one can obtain $\langle \sigma_i^A, \sigma_j^B \rangle$  by measuring the probe qubit thus circumventing the need for repeated direct measurements of the quantum systems of interest. 

The first spin $^{13}$C$_1$ (depicted in Fig.~\ref{molecular}) acts as the probe qubit, spins $^{13}$C$_{2,4}$ comprise the two quantum systems of interest, and $^{13}$C$_3$ constitutes the ancillary qubit, which intermediates the causal influence between the spins  $^{13}$C$_{2,4}$.

\begin{figure}[b]
\centering
\includegraphics[width=1\linewidth]{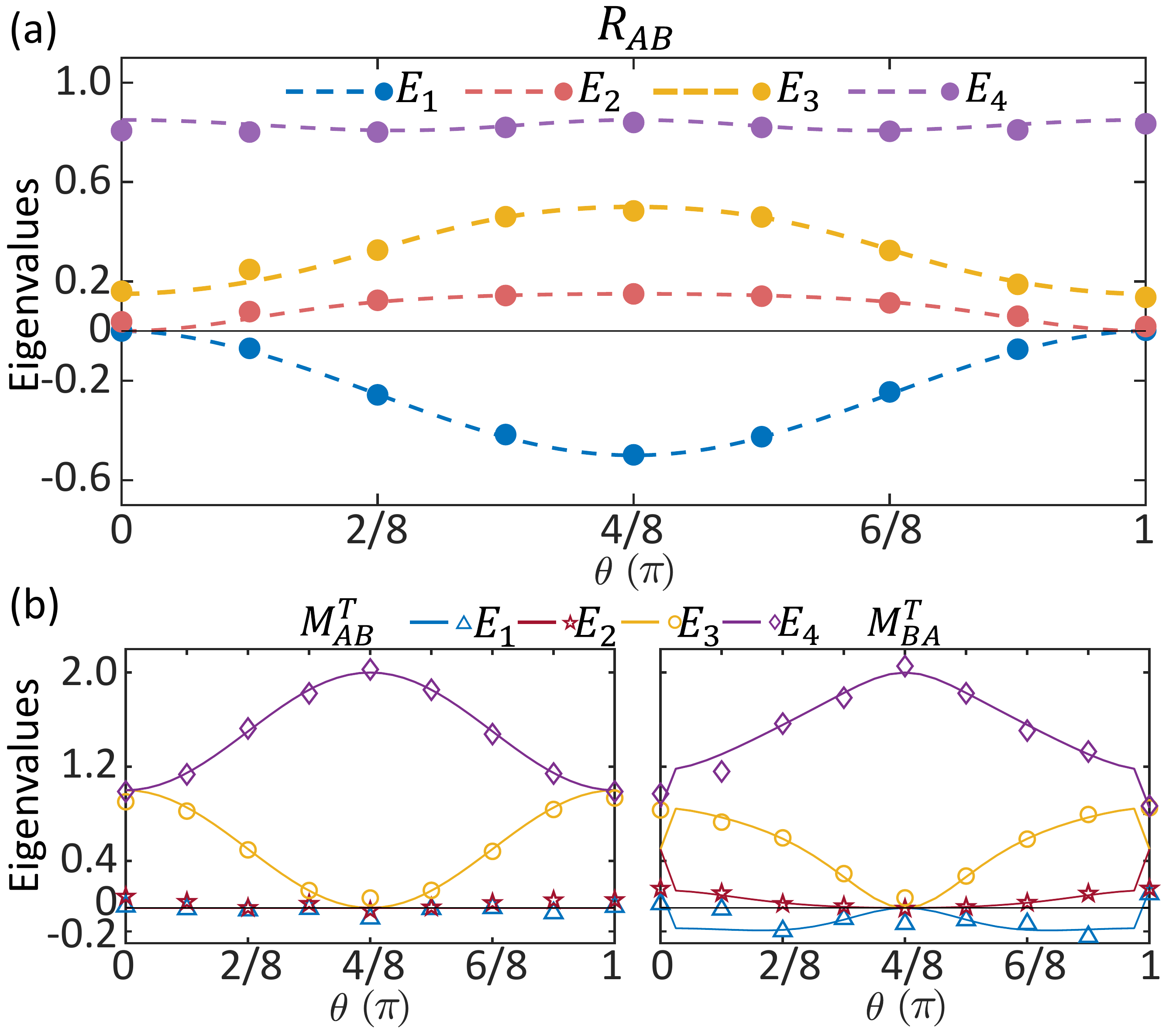}
\caption{{\bf Measured eigenvalues of $R_{AB}$, $M_{AB}^T$ and $M_{AB}^T$ for varying channel strength.} The input state is fixed at $\lambda=0.7$ and the channel strength $\theta$ varied from $0$ to $\pi$. The dashed line (solid line) represents the theoretical results, while the markers indicate the experimental results. The error bars are contained within the size of the markers.}
\label{SMresult1}
\end{figure}

\begin{figure}[t]
\centering
\includegraphics[width=1\linewidth]{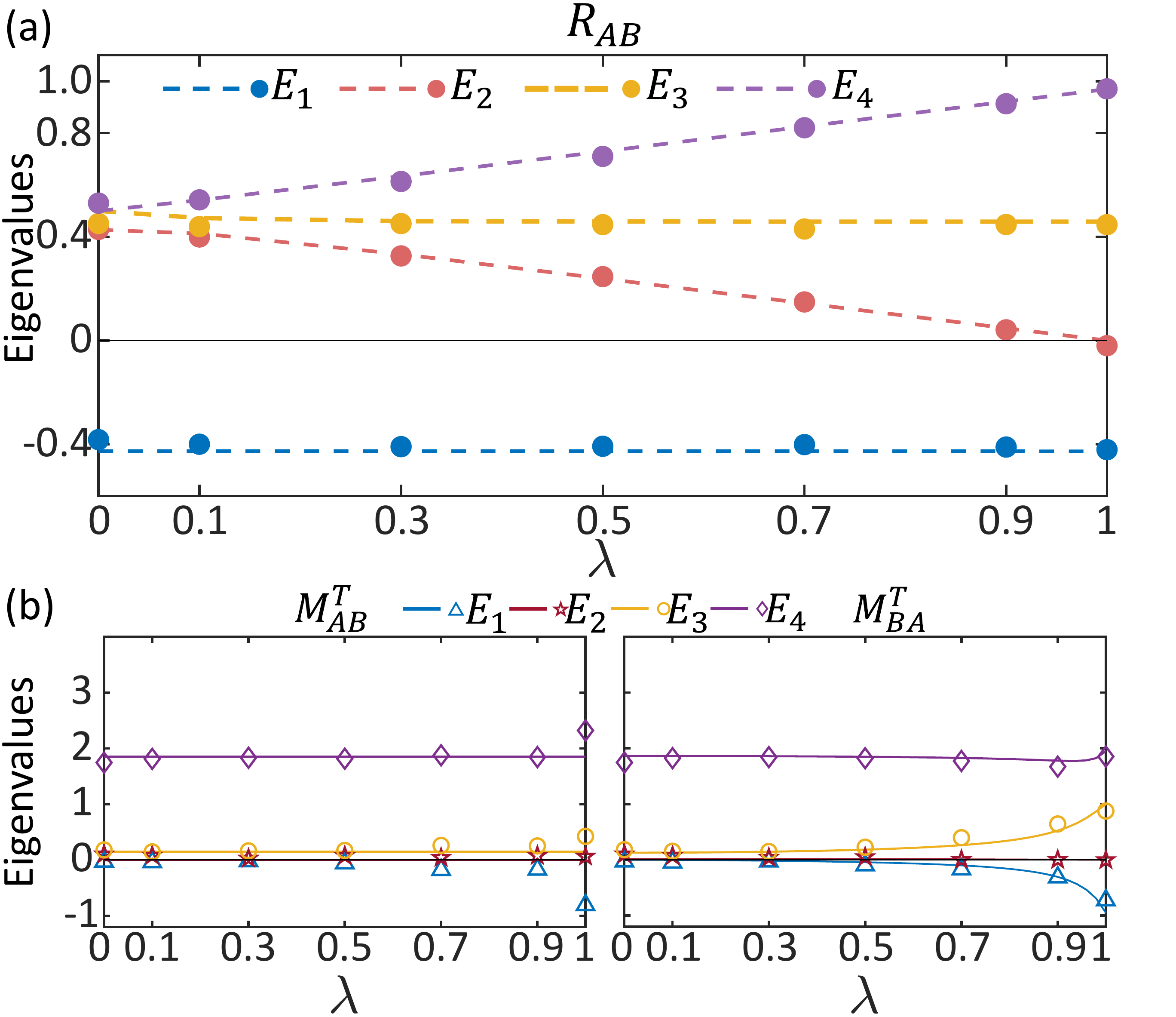}
\caption{{\bf Measured eigenvalues of $R_{AB}$, $M_{AB}^T$ and $M_{BA}^T$ for varying initial polarization.} The channel strength is fixed at $\theta=3\pi/8$ and the polarization of the initial state, $\lambda$, is varied from $0$ to $1$. The dashed line (solid line) represents the theoretical results, while the markers indicate the experimental results. The error bars are contained within the size of the markers}
\label{SMresult2}
\end{figure}

\textit{Unitary evolution.}--
The experimental circuit is designed to realize the unitary $\mathcal{U}_\text{c}$ $=\vert0\rangle\langle0\vert\otimes\mathbbm{I}+\vert1\rangle\langle1\vert\otimes \mathcal{V}\mathcal{W}$ on the four-qubit system.
The explicit forms of unitaries $\mathcal{V}$ and $\mathcal{W}$ are determined by the quantum channels, $\mathcal{N}$ and $\mathcal{M}$, as well as the time correlator $\langle \sigma_i^A, \sigma_l^B \rangle$ we aim to measure. 
Let $U_\mathcal{N}$ and $U_\mathcal{M}$ be the corresponding unitaries when $\mathcal{N}$ and $\mathcal{M}$ are unitary channels. In Fig.~\ref{circuit}, the explicit experimental quantum circuit is designed to obtain $\langle \sigma_i^A\sigma_j^C,\sigma_k^D\sigma_l^{B}\rangle$.    For this specific choice of the Pauli operators, $\mathcal{W}=(U_\mathcal{M}U_\mathcal{N})^\dagger\sigma_k\sigma_l(U_\mathcal{M}U_\mathcal{N})$ and $\mathcal{V}=\sigma_i\sigma_j$.
To obtain $\langle \sigma_i^A ,\sigma_l^B \rangle$, one set $\sigma^C=\sigma^D=\id$.
All the single-qubit rotations and two-qubit gates are fundamental quantum operations that can be easily implemented in most physical systems. For the NMR experimental apparatus, experimental control accuracy is further improved by creating a shaped pulse through gradient-based optimization~\cite{KHANEJA2005296}.

\textit{Results readout.}--
After the evolution mentioned above, the expectation value $\langle \sigma_i^A\sigma_j^C,\sigma_k^D\sigma_l^{B}\rangle$ is encoded in the coherence term of the probe qubit.  Subsequent application of a Hadamard gate allows direct extraction of the PDM element by measuring  $\langle \sigma_z\rangle$ of the probe qubit. In this way, we can construct the PDM $R_{ACDB}$, and the reduced PDM $R_{AB}$ when the expectation values  $\langle \sigma_i^A,\sigma_l^{B}\rangle$ are selected.
To construct $R_{AB}$ and $R_{ACDB}$, the experiment was repeated 16 and 256 times, respectively, each time with a refreshed $\mathcal{U}_\text{c}$ corresponding to the specific elements.

\textit{Experimental Results for Eigenvalues of Choi matrices}.--In our first experiment, we set $\lambda=0.7$ and systematically varied the channel strength $\theta$ over the range from $0$ to $\pi$, measuring the corresponding PDM $R_{AB}$. The data for the PDMs $R_{AB}$ is shown in Fig.~\ref{SMresult1}. We observed that the PDMs have negative eigenvalues, i.e., $f(R_{AB})>0$. Next, we calculated the Choi matrices of the process $M_{AB}^{T}$ and the corresponding time-reverse Choi matrices $M_{BA}^{T}$ from the experimental PDM $R_{AB}$. The eigenvalues of the Choi matrices allow us to deduce the cause and the effect. The experimental results of eigenvalues are shown in the Fig.~\ref{SMresult1} (b). For the matrices $M_{AB}^T$, as the channel strength $\theta$ varies, the eigenvalues of $M_{AB}^T$ remain largely positive. In contrast, for $M_{BA}^T$, negative eigenvalues can emerge. These results indicate that the data $R_{AB}$ is compatible with a cause-effect mechanism in which $A$ is the cause and $B$ is the effect (case $A\rightarrow B$ in Table~\ref{tab:protocol}), consistent with theoretical predictions.

In our second experiment, we varied the polarization parameter $\lambda$ while fixing the strength of the channel at $\theta=3\pi/8$ and measured the PDM $R_{AB}$. The data for the PDMs $R_{AB}$ is provided in Fig.~\ref{SMresult2} (a), and negative eigenvalues are observed. We proceed to calculate the experimental eigenvalues of the Choi matrices  $M_{AB}^T$ and $M_{BA}^T$, with results displayed in Fig.~\ref{SMresult2} (b). When the initial state is varied, different initial states pass through the same channel. Consequently, the corresponding matrix $M_{AB}^T$ computed using PDM $R_{AB}$ remains invariant, and the eigenvalues of $M_{AB}^T$ are constant and positive. However, for its time-reversed matrix
$M_{BA}^T$, different $M_{BA}^T$ are obtained. The magnitude of the negative eigenvalues of $M_{BA}^T$ increases with increasing polarization $\lambda$. These results too indicate that the data $R_{AB}$ is compatible with a cause-effect mechanism in which $A$ is the cause and $B$ is the effect (case $A\rightarrow B$ in Table~\ref{tab:protocol}), consistent with theoretical predictions.
\begin{figure}[t]
\centering
\includegraphics[width=1\linewidth]{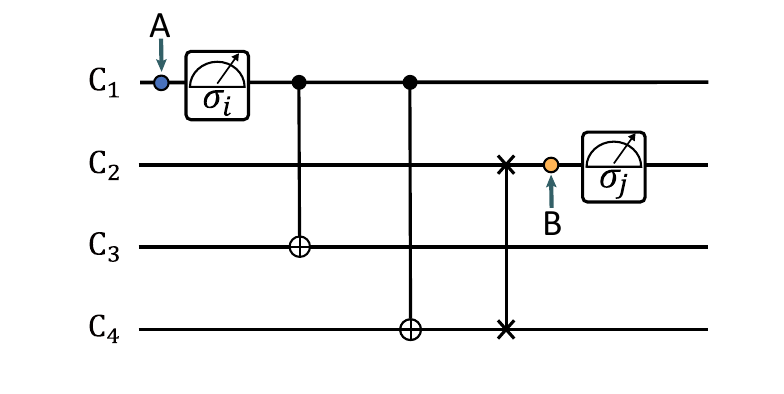}
\caption{{\bf Simulating the fully decohering channel and measuring $\langle \sigma_i^A, \sigma_j^B \rangle$.}  The first spin, $C_1$, carries the input state of the channel $\rho_A$, while the final state of the channel is output by the second qubit, $C_2$. The pseudo density matrix element $\langle \sigma_i^A, \sigma_j^B \rangle$ associated with two different systems and two different times is measured. }
\label{SMdpcircuit}
\end{figure}
\begin{figure*}[t]
\centering
\includegraphics[width=0.9\linewidth]{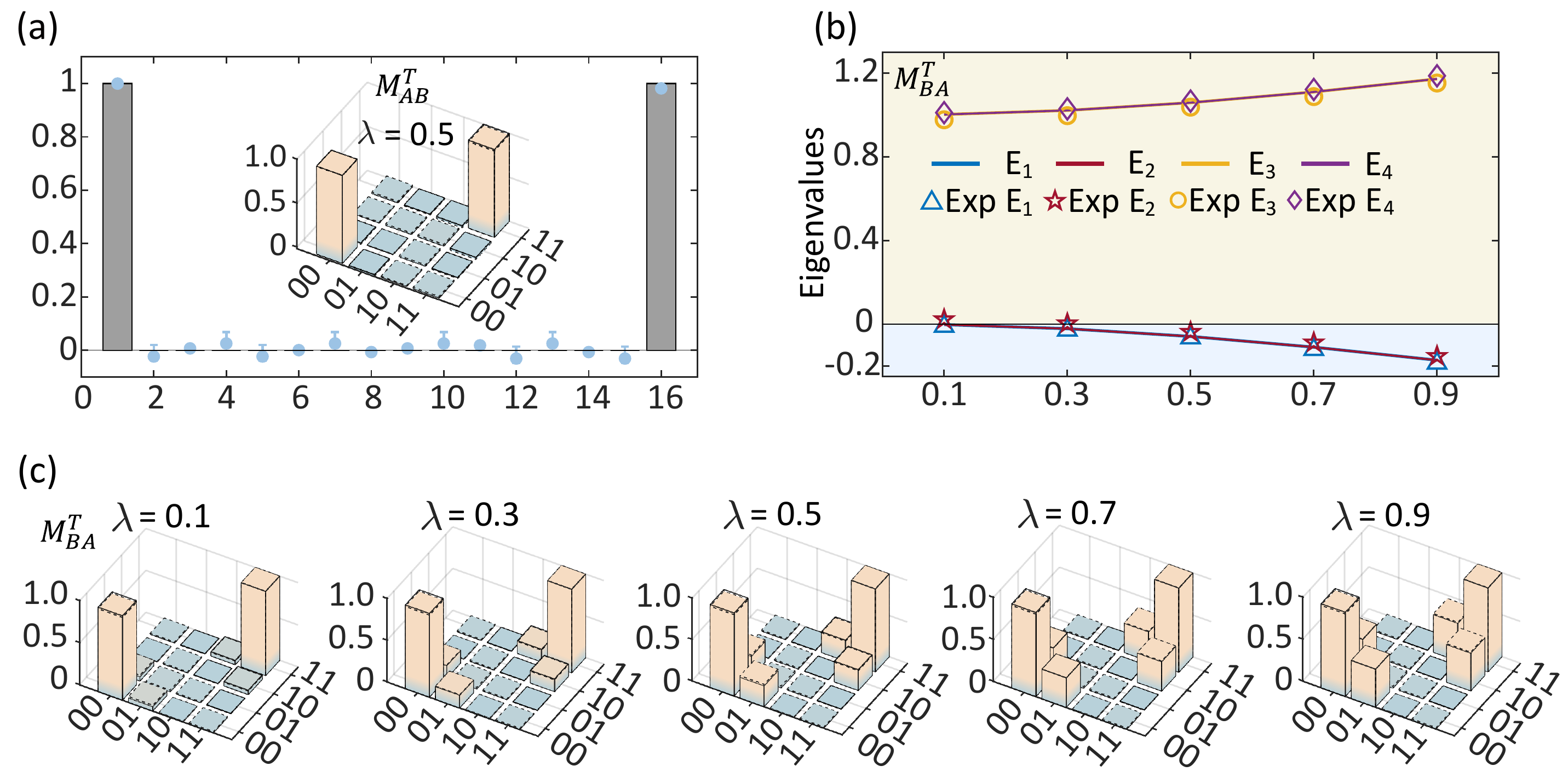}
\caption{{\bf Measured eigenvalues and Choi matrices.} 
(a) The experimental results of the Choi matrix $M^T_{AB}$ for all of the five $\lambda$. The $M^T_{AB}$ of the different initial states is the same. The circles represent the average value of five results, and the color bar represents the theoretical results. The bar form in the middle indicates the Choi matrix $M^T_{AB}$ when $\lambda=0.5$.
(b) The eigenvalues of $M_{BA}^T$ vary with $\lambda$. Markers indicate eigenvalue points obtained from the experiment, while the solid line depicts the corresponding theoretical results. (c) The Choi matrix $M^T_{BA}$ for different $\lambda$. The black solid line represents the theoretical results, while the color bars with dashed lines represent the experimental results.}
\label{SMresult3}
\end{figure*}

\section{Experimental implementation for structure with decohering channel}
We further extended our protocol to a fully decohering channel, demonstrating that the protocol is also effective for this evolution.

A quantum state $\rho_A$ passing through the fully decohering channel turns into 
\begin{equation}
    \mathcal{L}(\rho_{A})=\langle0|\rho_A\ket{0}\ket{0}\langle0|+\langle1|\rho_A\ket{1}\ket{1}\langle1|. 
    \label{FDC}
\end{equation}

\textit{Experimental scheme.}--We designed a circuit containing four qubits to simulate the quantum decoherence channel, as shown in Fig.~\ref{SMdpcircuit}. 
The first spin, C$_1$, carries the input state of the channel $\rho_A$, while the final state of the channel is output by the second qubit, C$_2$. The last two qubits, C$_3$ and C$_4$, are used to simulate the fully decoherent channel.
The initial state of the system is $\rho_i\otimes\vert0\rangle\langle0\vert\otimes\frac{\mathbb{I}}{2}\otimes\vert0\rangle\langle0\vert$, with $\rho_i=(1-\lambda)\frac{\mathbb{I}}{2}+\lambda\ket{+}\langle+|$.
This circuit contains three quantum gates. 
The first gate is a CNOT gate, with C$_1$ as the control qubit and C$_3$ as the target qubit. The density matrix of C$_1$  after the gate is
$$\inp{0|\rho_i|0}\op{0}{0}+\inp{1|\rho_i|1}\op{1}{1}.$$ 
The second gate is also a CNOT gate with C$_1$ as the control qubit and  C$_4$ as the target qubit. The joint state of   C$_1$ and  C$_4$ after this is
$$\inp{0|\rho_i|0}\op{00}{00}+\inp{1|\rho_i|1}\op{11}{11}.$$ 
The third gate is a swap operation. This operation replaces the quantum state of C$_2$ with C$_4$. Therefore,
the joint state of C$_1$ and C$_3$ after the three gates is given by $$ \langle0|\rho_i\ket{0}\ket{00}\langle 00|+\langle 1|\rho_i\ket{1} \ket{11}\langle 11|.$$ 
When looking at the input state $\rho_i$ of C$_1$ and the final state, $\langle0|\rho_i\ket{0}\ket{0}\langle 0|+\langle 1|\rho_i\ket{1} \ket{1}\bra{1}$, of C$_2$, our circuit presents a successful simulation of the fully decohering channel.

Due to the limited number of qubits of the quantum simulator in our experiments, we could not use the scattering circuit to measure the PDM. 
In the experiment, we first prepared spin C$_1$ in the state $\rho_A$ and then measured the probabilities of its eigenstates under different Pauli operators $\sigma_i$. Subsequently, $\rho_A$ is collapsed into a specific eigenstate and passed through the designed circuit. Finally, we perform quantum state tomography on the final state of C$_2$ to calculate the probabilities of the eigenstates under various Pauli operators $\sigma_i$. These data are then used to compute the expectation values $\langle \sigma_i^A, \sigma_j^B \rangle$ according to the method described in Ref \cite{fitzsimons2015quantum}, thereby obtaining the PDM $R_{AB}$. The PDM can also be measured using a scattering circuit on a five-qubit quantum simulator.

We studied the fully decohering channel for different input states, varying $\lambda$ from $0$ to $1$. The Choi matrix $M_{AB}^T$ remains the same and is positive for different values of $\lambda$, as shown in Fig.~\ref{SMresult3} (a). The thus obtained time-reversed Choi matrices $M_{BA}^T$ based on experimentally constructed $R_{AB}$ vary with $\lambda$.   The eigenvalue results for the $M_{BA}^T$ and the bar form of $M_{BA}^T$ corresponding to different $\lambda$ are illustrated in Fig.~\ref{SMresult3} (c). The eigenvalues of  $M_{BA}^T$ as a function of $\lambda$ are plotted in Fig.~\ref{SMresult3} (b). Based on our causal inference protocol in Table~\ref{tab:protocol},  the causal structure is thus inferred to be compatible with $A \rightarrow B$.

\section{Experimental Error in $R_{AB}$}

\begin{figure}[t]
\centering
\includegraphics[width=1\linewidth]{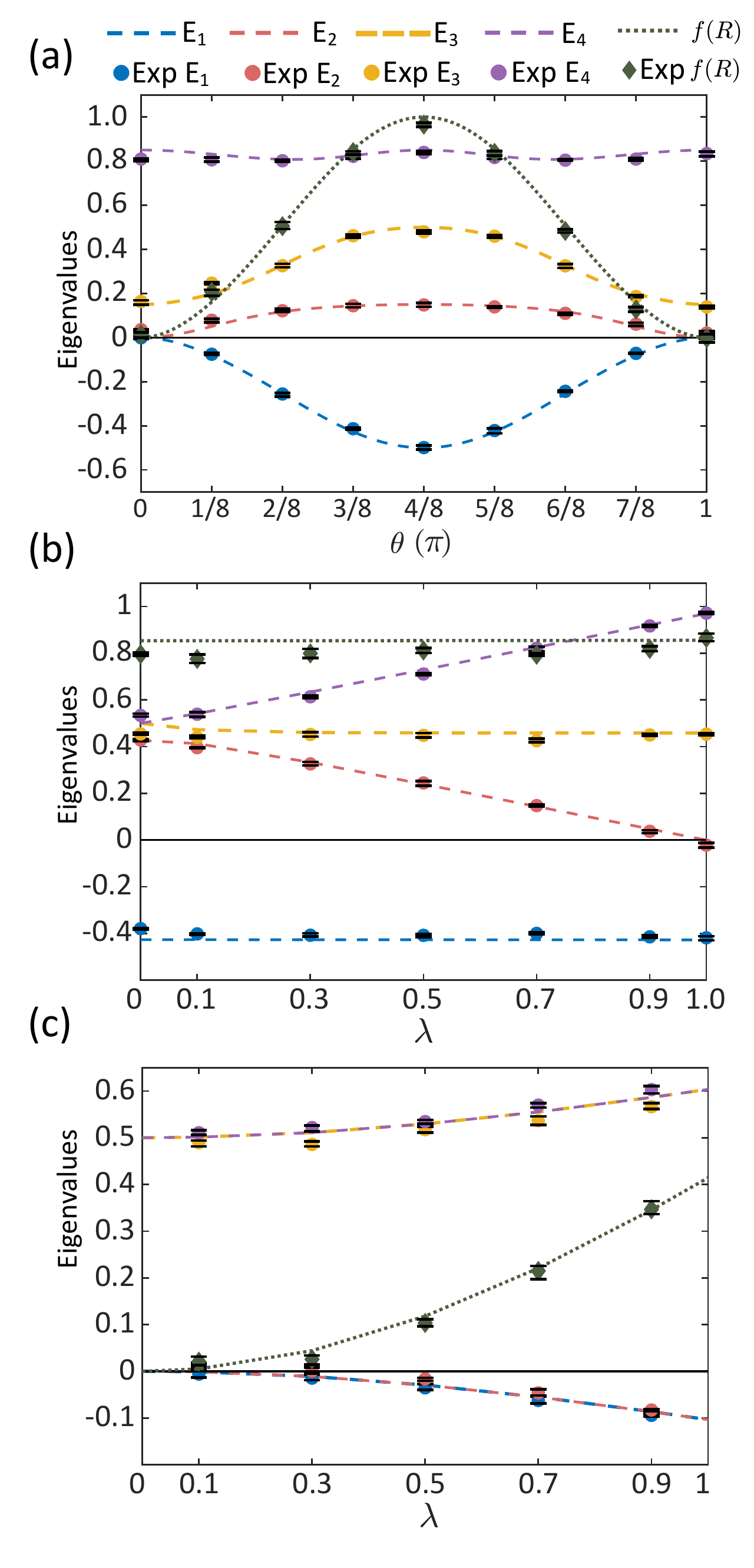}
\caption{{\bf Error bars of eigenvalues are within marker sizes.} The eigenvalues E$_i$ and $f(R_{AB})$ of $R_{AB}$ in the main text with error bars derived from five experimental repetitions. (a-b) The $R_{AB}$ for different parameters obtained from the partial swap channel. (c) The $R_{AB}$ for different $\lambda$ obtained from the fully decohering channel.}
\label{Error}
\end{figure}

In NMR platforms, the experimental samples do not consist of a single molecule, but rather a system composed of a thermodynamic limit number of identical molecules.
Consequently, the quantum state measurements performed in NMR are measurements of ensemble averages, which inherently stabilize against considerable fluctuations. Moreover, our experimental duration is controlled to remain within the decoherence time. Thus the effects of decoherence are negligible, leading to highly reliable results. The experimental error primarily arises from imperfections in quantum control.
We have incorporated error bars in all $R_{AB}$ mentioned in the main text, derived from five experimental repetitions, as illustrated in Fig.~\ref{Error}. Each experimental result is almost constant, resulting in error bars that fall within the size of the markers.

\section{Extracting $M$ when the initial quantum state is rank-deficient}
\label{sec:SDP}
When $\rho_A =\tr_{B}R_{AB}$ is rank-deficient, this section provides an algorithm for extracting the CJ matrix $M_{AB}$ from 

\begin{align}
    R_{AB}=\frac{1}{2}(\rho_A \otimes \mathbb{I}_B \, M_{AB} +M_{AB}  \, \rho_A \otimes \mathbb{I}_B),
\end{align}
for the corresponding Choi matrix $M^{T_A}_{AB}$ being the least negative.

 $M_{AB}$ can be uniquely extracted via the vectorization method in the full rank case~\cite{liu2023inferring,liu2023quantum}, whereas $M_{AB}$ is not unique for the case of $\rho_A$ being rank deficient. As proposed in Ref.~\cite{liu2023quantum}, the following semi-definite programming (SDP) is proposed to give a CJ matrix $M_{AB}$ with its Choi matrix $M^{T_A}_{AB}$ being the least negative,
\begin{equation}
\begin{aligned}
    & \text{minimize } \tr M_{-}^{T_{A}}
    \\
    & \text{subject to }
    \left\{\begin{aligned}
     & M = M^{\dagger}, \\
     & R = (\rho\otimes \mathbb{I}) M +M (\rho\otimes \mathbb{I}), \\
     & M^{T_{A}}=M^{T_{A}}_{+}-M^{T_{A}}_{-}, \\
     & \tr_{B} M = \mathbb{I}_{B}, \\
     & M^{T_{A}}_{+} \geq 0 , \quad M^{T_{A}}_{-} \geq 0.
    \end{aligned} \right.
\end{aligned}
\end{equation}
This SDP can be solved numerically with any computer capable of implementing convex optimization and partial transpose. There are various options. Here, we establish the SDP in Matlab and apply the Packages {\tt{CVX}} (for convex optimization) and {\tt{QETLAB}} (for implementing partial transpose).
The code for the SDP problem is then provided by the function {\tt{causalinfer}} as follows:
\begin{lstlisting}[
frame=single,
style=Matlab-Pyglike]
function N = causalinfer(R,d)
rho=kron(PartialTrace(R,2,[d,d]),eye(d))
cvx_begin sdp
variable Np(d*d,d*d) hermitian
variable Nn(d*d,d*d) hermitian
N=Np-Nn;
Nptp=PartialTranspose(Np,1,[d,d]);
Nntp=PartialTranspose(Nn,1,[d,d]);
minimize( trace(Nntp) )
0.5*(rho*N+N*rho)==R;
PartialTrace(N,2,[d,d])==eye(d);
Nptp >= 0;
Nntp >= 0;
cvx_end
N=N+zeros(d*d);
end
\end{lstlisting}
The output of the SDP is thus obtained by typing {\tt{causalinfer(R,d)}} where $R$ is the PDM and $d$ the dimension of $R$ in the sense that $R$ is $d\times d$.

\section{Conclusion and Discussion}
Our experimental results demonstrate the feasibility of inferring causal structures in quantum systems using coarse-grained projective measurements via scattering circuits. The experiments involving partial swap channels and fully decohering channels affirmed the applicability of this measurement-based approach. Notably, the successful inference of the causal structure from the fully decohering channel, where coherence was entirely lost, demonstrates that causal information can still be extracted even without coherence.

The ability to infer causal relationships even from fully decohering channels broadens the applicability of these techniques to a wider range of physical processes, including those where coherence is limited or absent. This suggests that causal inference based on measurements is more versatile than previously thought, potentially extending to diverse quantum scenarios. The use of negativity of the PDM, along with the evaluation of time asymmetry, proved effective in distinguishing different causal structures, highlighting their role as key indicators in identifying causal relationships within quantum systems. The approach may lead to new types of multiqubit channel tomography protocols based on coarse-grained measurements alone.

{\bf {\em Acknowledgements.---}} This work is supported by the National Key Research and Development Program of China (2019YFA0308100), and the National Natural Science Foundation of China (Grants No.12104213, 12075110, 12204230,
 12050410246, 1200509, 12050410245), Science, Technology and Innovation Commission of Shenzhen Municipal
ity (JCYJ20200109140803865), Guangdong Innovative and Entrepreneurial Research Team Program (2019ZT08C044), Guangdong Provincial Key Laboratory (2019B121203002), City University of Hong Kong (Project No. 9610623), The Pearl River Talent Recruitment Program (2019QN01X298), Guangdong Provincial Quantum Science Strategic Initiative (GDZX2303001 and GDZX2200001), and the National Research Foundation, Prime Minister’s Office, Singapore under its Campus for Research Excellence and Technological Enterprise (CREATE) programme. QC is funded within the QuantERA II Programme that has received funding from the European Union’s Horizon 2020 research and innovation programme under Grant Agreement No 101017733 (VERIqTAS). YQ is supported by the National Research Foundation, Singapore, and
A*STAR under its CQT Bridging Grant and its Quantum Engineering Programme under grant NRF2021-
QEP2-02-P05. V.V is supported by the Gordon and Betty Moore and Templeton Foundations.

\bibliography{references.bib}

\end{document}